\documentclass[a4paper]{jpconf}
\usepackage{graphicx}
\usepackage{epstopdf}
\usepackage{color}

\begin{document}
\title{New insights on the solar core}
\author{R.A. Garc\'\i a$^1$, D. Salabert$^{2,3}$, J. Ballot$^4$, A. Eff-Darwich$^{2,5}$, R. Garrido$^6$, A. Jim\'enez$^{2,3}$, S. Mathis,$^{1}$, S. Mathur$^{7}$, A. Moya$^8$, P.L. Pall\'e$^{2,3}$, C. R\'egulo$^{2,3}$, K. Sato$^1$, J.C. Su\'arez$^6$ and S. Turck-Chi\`eze$^{1}$}
\address{$^1$ Laboratoire AIM, CEA/DSM-CNRS, Universit\'e Paris 7 Diderot, IRFU/SAp, Centre de Saclay, 91191, Gif-sur-Yvette, France}
\address{$^2$ Instituto de Astrof\'isica de Canarias (IAC), E-38200 La Laguna, Tenerife, Spain}
\address{$^3$ Dept. de Astrof\'isica, Universidad de La Laguna (ULL), E-38206 La Laguna, Tenerife, Spain}
\address{$^4$ Laboratoire d'Astrophysique Toulouse-Tarbes - OMP, 14 avenue Edouard Belin, 31400 Toulouse, France}
\address{$^5$ Dept. de Geolog\'\i a, Universidad de La Laguna (ULL), E-38206 La Laguna, Tenerife, Spain}
\address{$^6$ Instituto de Astrof\'\i sica de Andaluc\'\i a (CSIC), Apartado 3004, 18080 Granada, Spain}
\address{$^7$ High Altitude Observatory, 3080 Center Green Drive, Boulder, CO, 80302, USA}
\address{$^8$ Centro de Astrobiolog\'\i a (INTA-CSIC), Madrid, Spain}

\ead{rgarcia@cea.fr}




\begin{abstract}	
Since the detection of the asymptotic properties of the dipole gravity modes in the Sun, the quest to find individual gravity modes has continued. An extensive and deeper analysis of 14 years of continuous GOLF/SoHO observational data, unveils the presence of a pattern of peaks that could be interpreted as individual dipole gravity modes in the frequency range between 60 and 140 microHz, with amplitudes compatible with the latest theoretical predictions. By collapsing the power spectrum we have obtained a quite constant splitting for these patterns in comparison to regions where no g modes were expected. Moreover, the same technique applied to simultaneous VIRGO/SoHO data unveils some common signals between the power spectra of both instruments. Thus, we are able to identify and characterize individual g modes with their central frequencies, amplitudes and splittings allowing to do seismic inversions of the rotation profile inside the solar core. These results open a new ligh
 t on the physics and dynamics of the solar deep core.
\end{abstract}

\section{Introduction}
Little progress has been done during the last few years on the structure [1,2,3] and Dynamics [4,5,6,7] of the solar core even after the detection of the asymptotic spacing of the dipole gravity modes [8,9] Indeed no general consensus has been obtained for the detection of individual g modes yet [10].  This is because the increasing convective background level towards the lower frequencies (e.g. [11]), combined with very small amplitudes of those modes (several mm/s in the case of g modes, [12]) are the limiting factors for their detections. In the case of the low-degree, low-frequency p modes, accurate measurements are hardly obtained below 1~mHz [13,14,15].

This situation might change in the near future when we have a few years of data coming from the new instrumentation available, such as PICARD [16], and the very promising HMI and AIA aboard SDO [17] or the new projected instrumentation (e.g. GOLF-NG [18]).

Fortunately, time goes by and the signal-to-noise ratio of the Global Oscillations at Low Frequency (GOLF) instrument [19] and the Sun Photometers (SPM) on the Variability of IRradiance and Global Oscillations [20] aboard the Solar and Heliospheric Observatory (SoHO) mission increases. In this work we uncover the presence of peaks in the power spectral density (PSD) that could be individual $\ell$=1 modes. These peaks are regularly spaced in period in the positions determined by the asymptotic periodicity measured by [8] Moreover, these peaks present a regular pattern in frequency that could be the signature of the rotational splitting. Thus this work study potential candidates that might merit further investigation, a step forward of what it has been done during the last few years (e.g. [21,22,23]).

\section{Observations and Data analysis}
In this work, we have used two different sets of data starting on April 11, 1996. The first one contains 4472 days and the second 5163 days finishing on May 30, 2010. The GOLF signal has been calibrated into velocity [24] following the methods described in [25,26].
We have worked with a single -- full resolution -- power spectrum even knowing that GOLF has been observing in two different configurations (blue and red) with a different sensitivity to the visible solar disk [27,28]. Because the analysis we are going to do is very sensitive to the amplitude of the peaks, we have computed a 5 times zero-padded power spectrum to reduce the problem of the discretization in frequency (see the discussion concerning this problem in [29]). We have also used time series from the SPM/VIRGO package on the same time span.

To uncover the peaks in the PSD responsible for the measurement of the periodic signal found in GOLF by [8] and interpreted as the spacing of the dipole gravity modes, we follow the same procedure usually done in low signal-to-noise ration (SNR) targets in asteroseismology. Thanks to the observations of solar-like stars done by CoRoT [30] and Kepler [31] and some ground based campaigns [32] we now know how to tackle the problem and measure individual acoustic (p) modes in such stars [33,34,35,36]  without ambiguity. Even when the SNR of these modes is very small, we are still able to determine the large separation even without seeing the individual p modes in the power spectrum (e.g. [37,38,39,40,41,42,43]). In such cases, the only thing that is generally done is to perform a more or less heavy smooth of the power spectrum to unveil the p-mode hump in the region in which the frequency spacing is found. 

\section{Results}

\begin{figure}[!htb]
\includegraphics[trim = 1.2cm 3cm 1cm 1cm, width = 0.8\textwidth, angle=270]{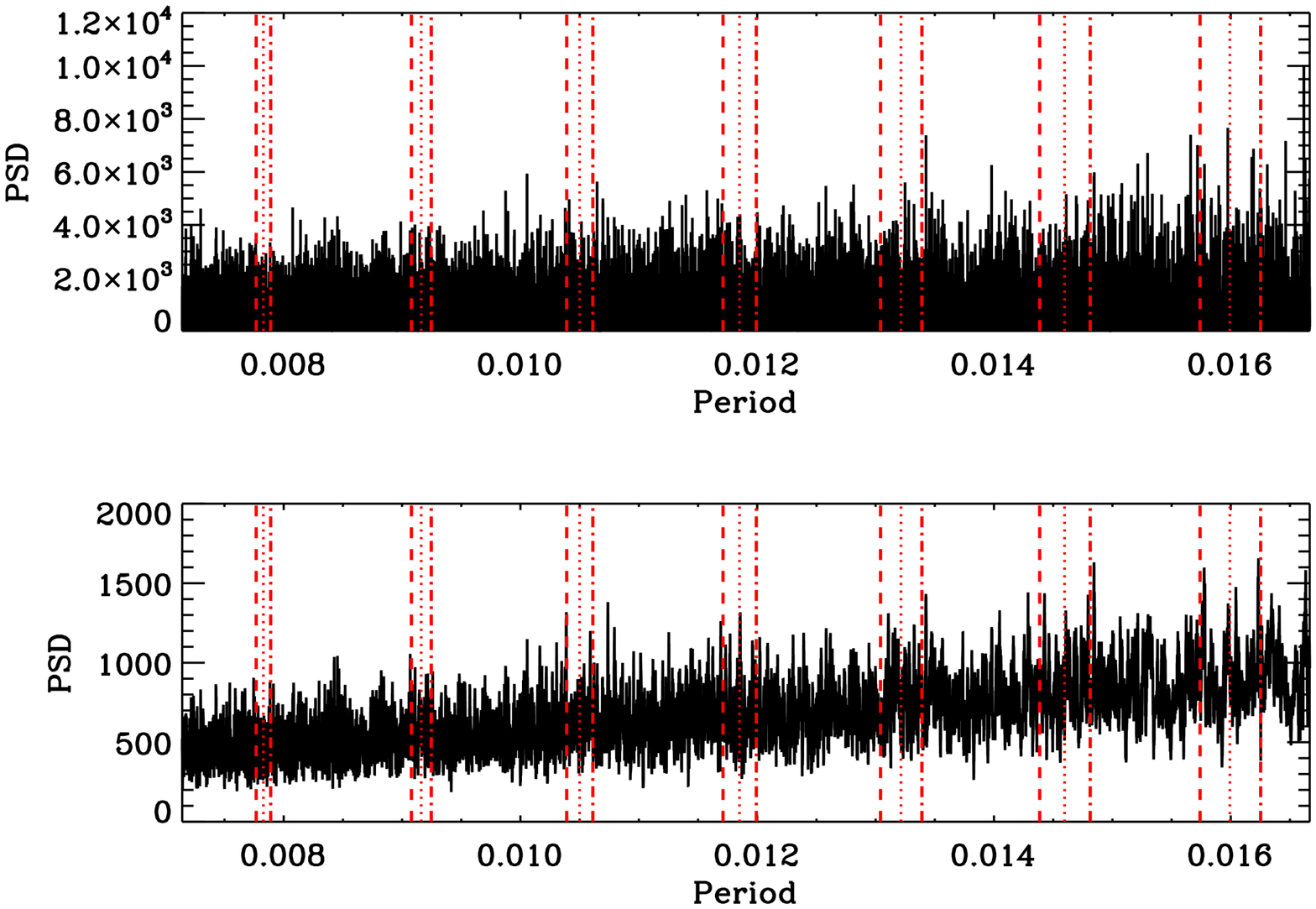}	
\caption{PSD in $(m/s)^2/Hz$ of the zero padded series (top) and the one after applying a smooth by a boxcar of 41 nHz (bottom), as a function of the period expressed in Mega seconds (inverse of $\mu$Hz) in the region [60-140] $\mu$Hz. The vertical dotted lines are the $\ell$=1 predicted frequencies. The dashed and dot-dashed lines represent the positions of the $m=\pm1$ components assuming a rotational splittings of 2 $\mu$Hz.
\label{fig:GOLF}}
\end{figure}

As said previously, we have computed the 5 times zero-padded Power Spectral Density (PSD) of the GOLF time series, which is represented in Figure~\ref{fig:GOLF} (top) as function of the period. Then, we took the predicted frequencies of the dipole gravity modes $\ell$=1, $m$=0 components from the Saclay seismic model [44,2] and we overplotted them to guide the eye of the reader (vertical red dotted lines).  We have also checked that in this frequency range of [60-140] $\mu$Hz, the differences between the frequencies of this model and the ones of the model S [45] and the model M1 from Nice [46] are less than 0.05 $\mu$Hz --which is inside the width of the vertical lines--. As the frequencies of the modes can be dependent of the pulsation code used to compute them, we have also verified how the predictions of the $\ell$=1 in the same frequency range vary with the parameters of the pulsation codes: the differences are negligible [47].

The dashed and dot-dashed vertical red lines are the m=$\pm$1 assuming a core rotating 5 times faster than the rest of the radiative region (2~$\mu$Hz). This value is compatible with the range inferred by [8] for the core rotation rate  and it matches the highest peaks of the PSD. Then, as the SNR is quite low, we smoothed the PSD with a boxcar function of 41 nHz (see Figure~\ref{fig:GOLF}, bottom). We can see that, in most of the cases, the high-amplitude peaks are located around the $m=\pm1$ components. It is important to notice that similar results are obtained using different sizes of the boxcar filter.

\begin{figure*}[h*]
\centering
\begin{tabular}{cc}
\includegraphics[angle=90, width = 0.55\textwidth, angle=270, trim = 1cm 2cm 1cm 1cm, clip]{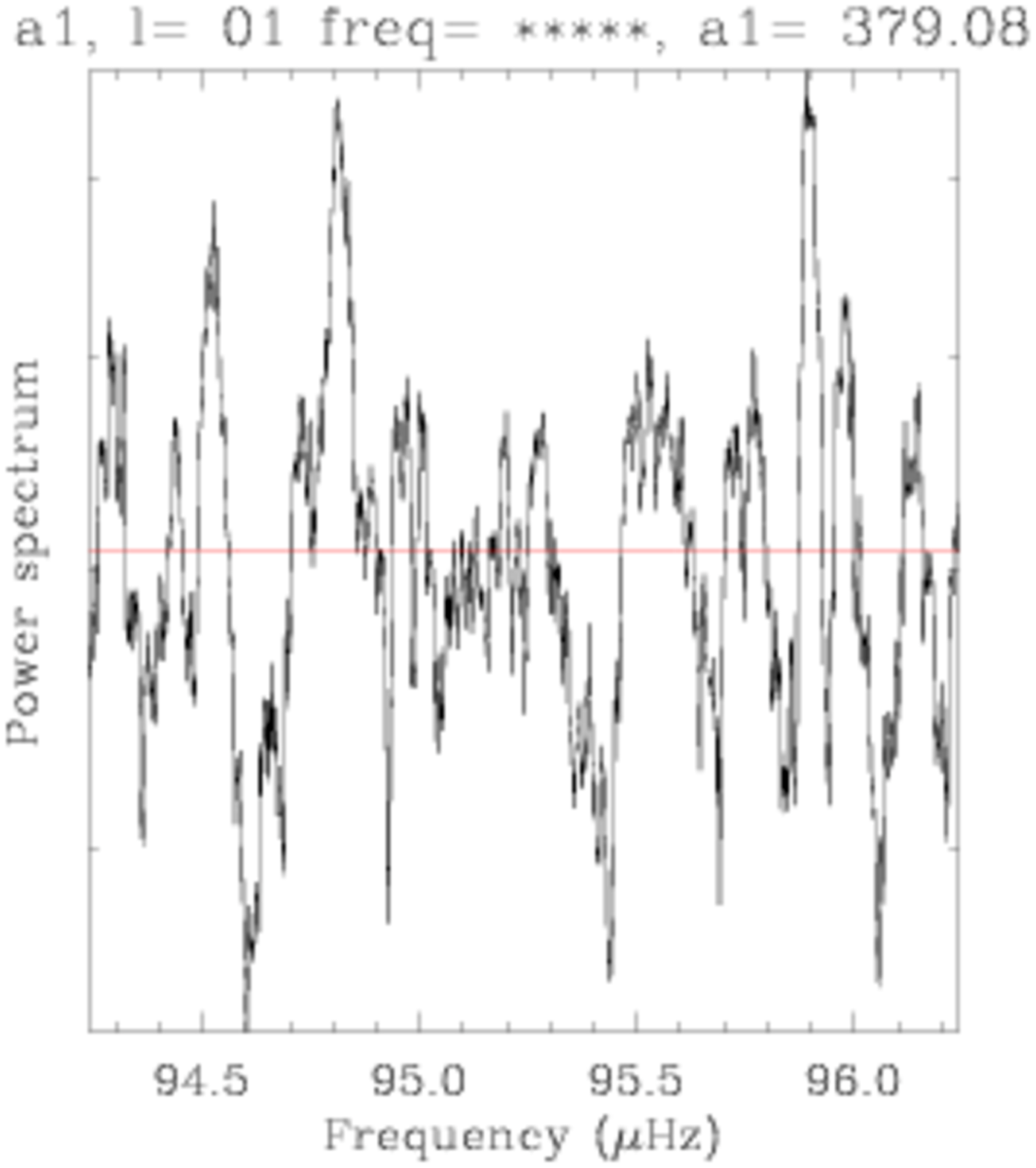}&  	
\includegraphics[angle=90, width = 0.55\textwidth, angle=270,trim = 1cm 2cm 1cm 1cm,clip]{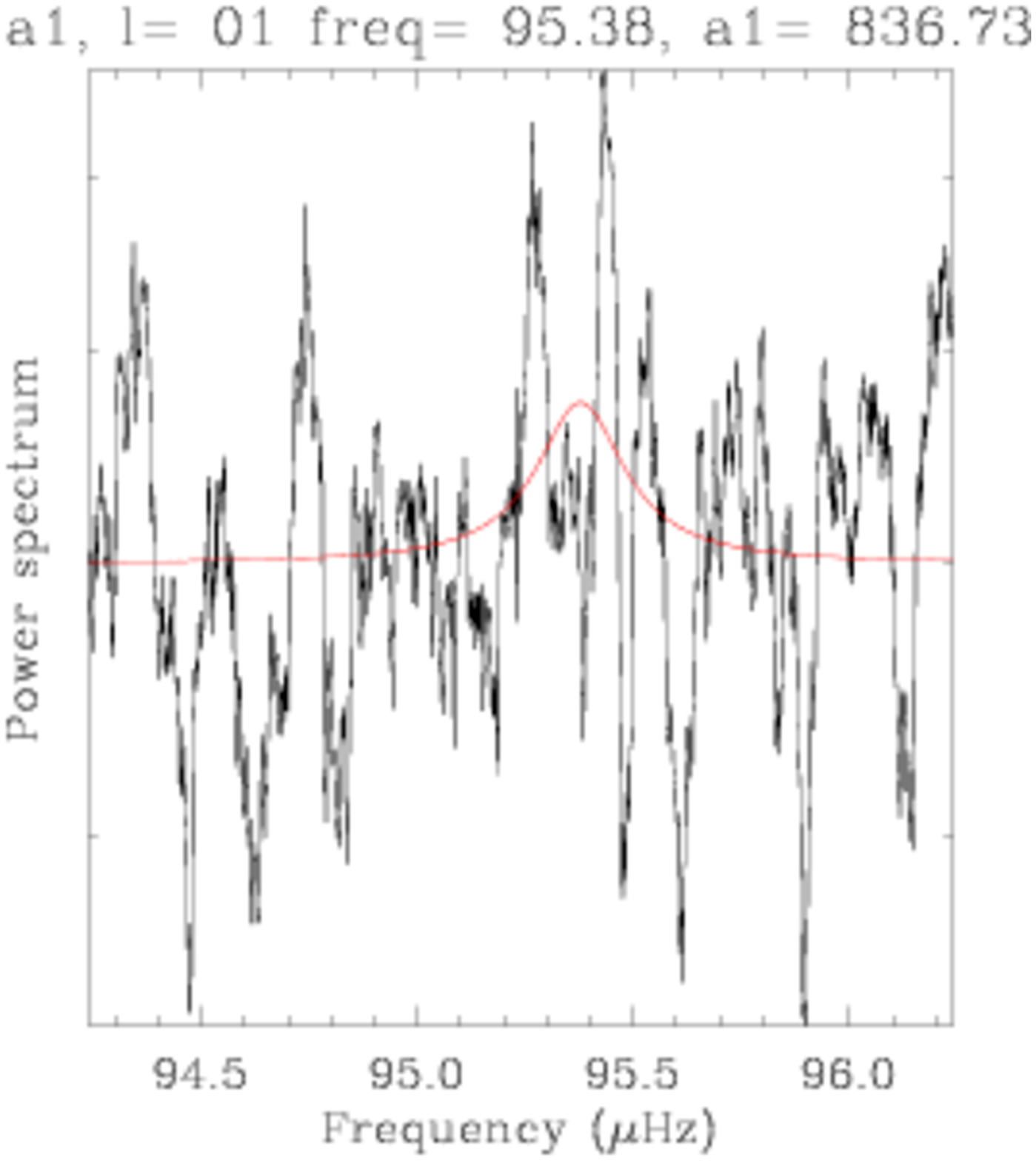}\\	  	
\includegraphics[angle=90, width = 0.55\textwidth, angle=270,trim = 1cm 2cm 1cm 1cm,clip]{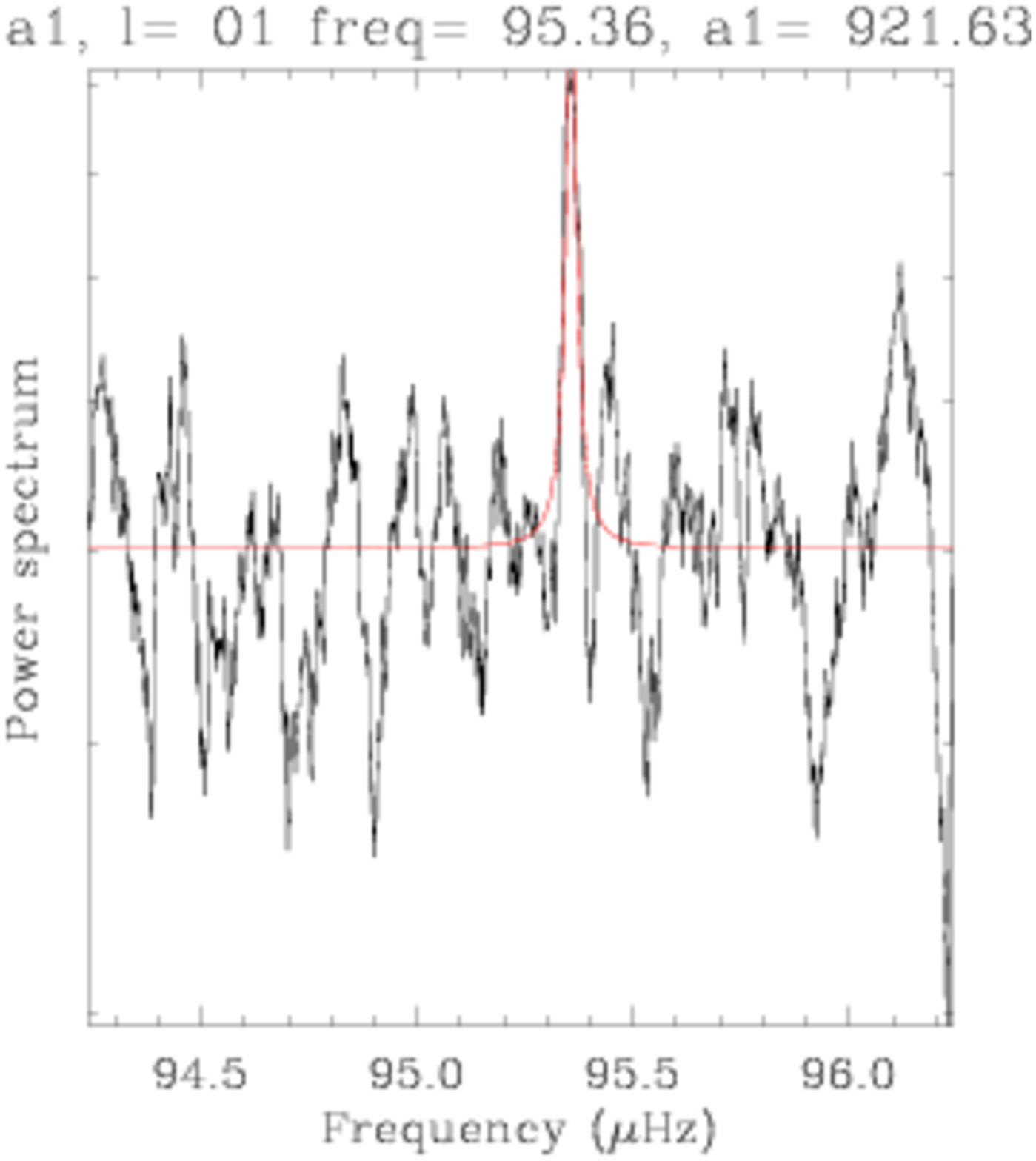}& 	
\includegraphics[angle=90, width = 0.55\textwidth, angle=270,trim = 1cm 2cm 1cm 1cm,clip]{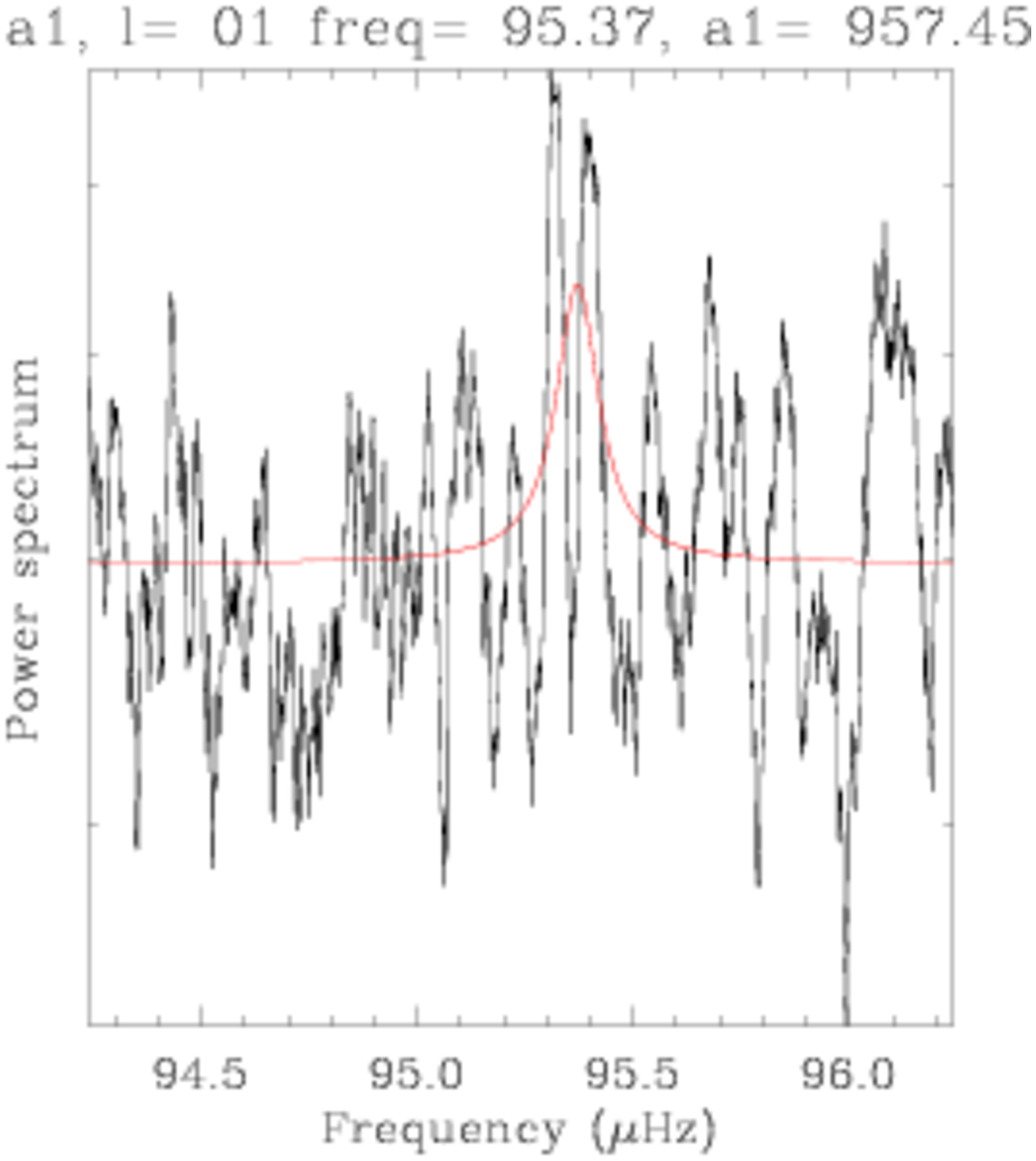} 	
\end{tabular}
\caption{Examples of the $m$-averaged spectrum technique applied to the g mode $\ell$=1, n=-6. The collapsograms are shown for four different rotational splittings at 379.08, 836.73, 934.90, and 957.45~nHz. The red solid line represents the fitted Lorentzian function. The fitted central frequency (when successfully fitted) is also given (in $\mu$Hz). The top-left panel shows an example in which no Lorentzian function were fitted (labeled with asterisks in the title).
\label{collapso}}
\end{figure*}

To quantitatively measure these g-mode candidates, we used the rotation-corrected, $m$-averaged spectrum technique, the so-called collapsograms. This method was originally developed for imaged instruments (see [15] for a detailed explanation of the method) to observe low SNR, low-frequency solar p modes. An $m$-averaged spectrum corresponds to the average of the individual $m$ components, thus reducing the non-coherent noise. Before averaging, each $m$ spectrum of a given mode ($n,l$) is shifted by a frequency that compensates for the effect of rotation and structural effects on the frequencies. The $m$-averaged spectrum concentrates for a given multiplet ($n,l$) all the $m$ components, considerably improving the SNR of the resulting spectrum, which is then fitted using a Lorentzian profile. The shift coefficients are chosen to maximize the likelihood of the $m$-averaged spectra. Figure \ref{collapso} shows a sequence of four snapshots of the procedure, for four different shifts (379.08, 836.73, 934.90, and 957.45~nHz respectively from left to right and top to bottom), around the theoretical frequency of the g mode $\ell$=1, $n=-6$. The central frequency of the mode is given by the frequency of the fitted Lorentzian profile, while the displacement of the $m$-components provides a measurement of the rotational splitting.

In five out of the seven dipole gravity modes in the region [60-140] $\mu$Hz we were able to obtain a stable splittings (around 850 to 950 nHz) around a central frequency laying close to the predicted frequencies. The modes we found with this method were the $\ell$=1, n=[-4,-6,-8,-9,-10]. 

The same work has been performed using the VIRGO/SPM datasets. Although the PSD is noisier than the GOLF one, in some cases we obtained a reliable Lorentzian fitting giving the same quantitative results in the central frequency of the mode and the splittings compared to GOLF.  

\section{Dynamics of the solar core}

\begin{figure}[!htb]
\includegraphics[angle=90, width = 0.35\textwidth, angle=270]{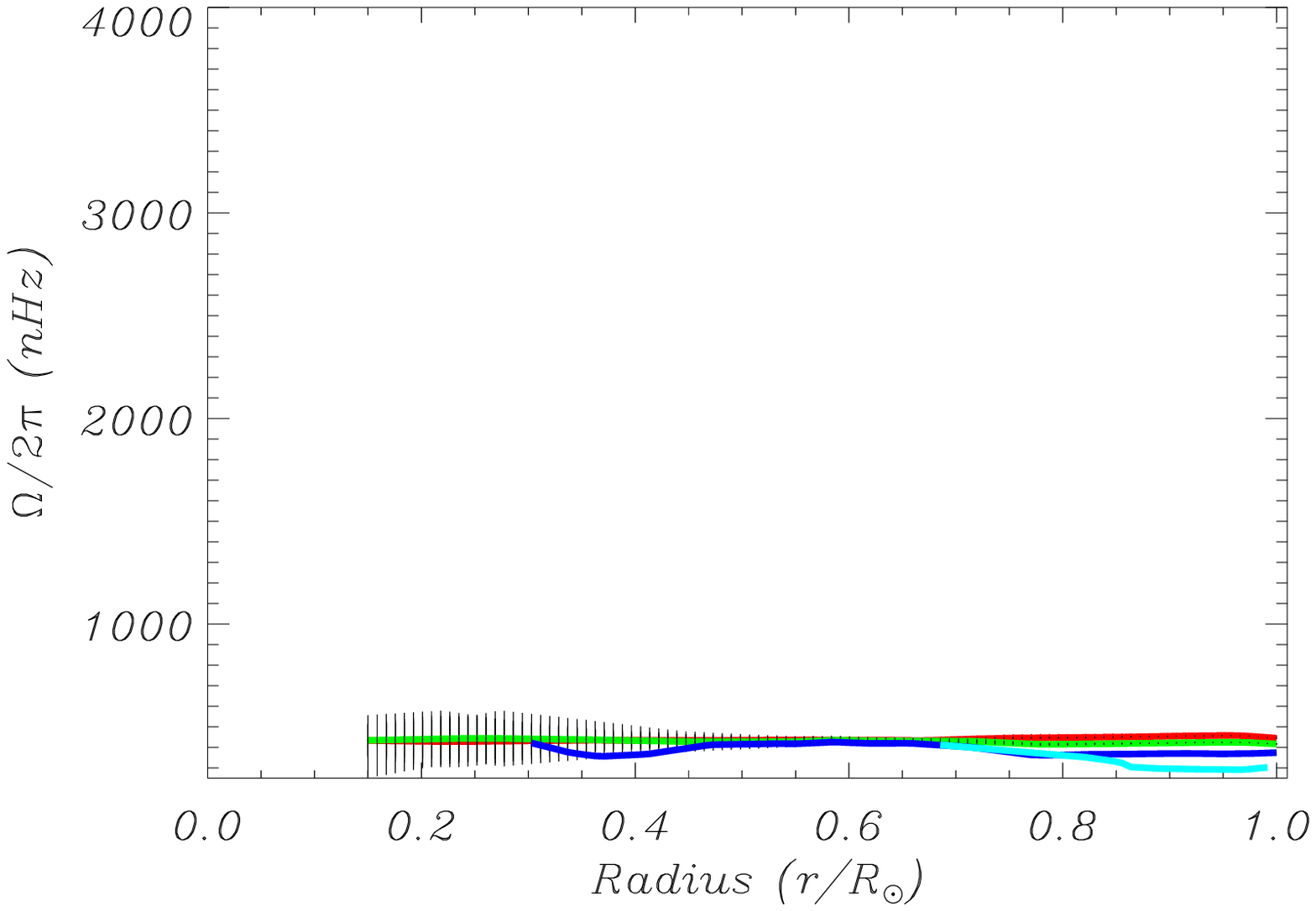}	
\includegraphics[angle=90, width = 0.35\textwidth, angle=270]{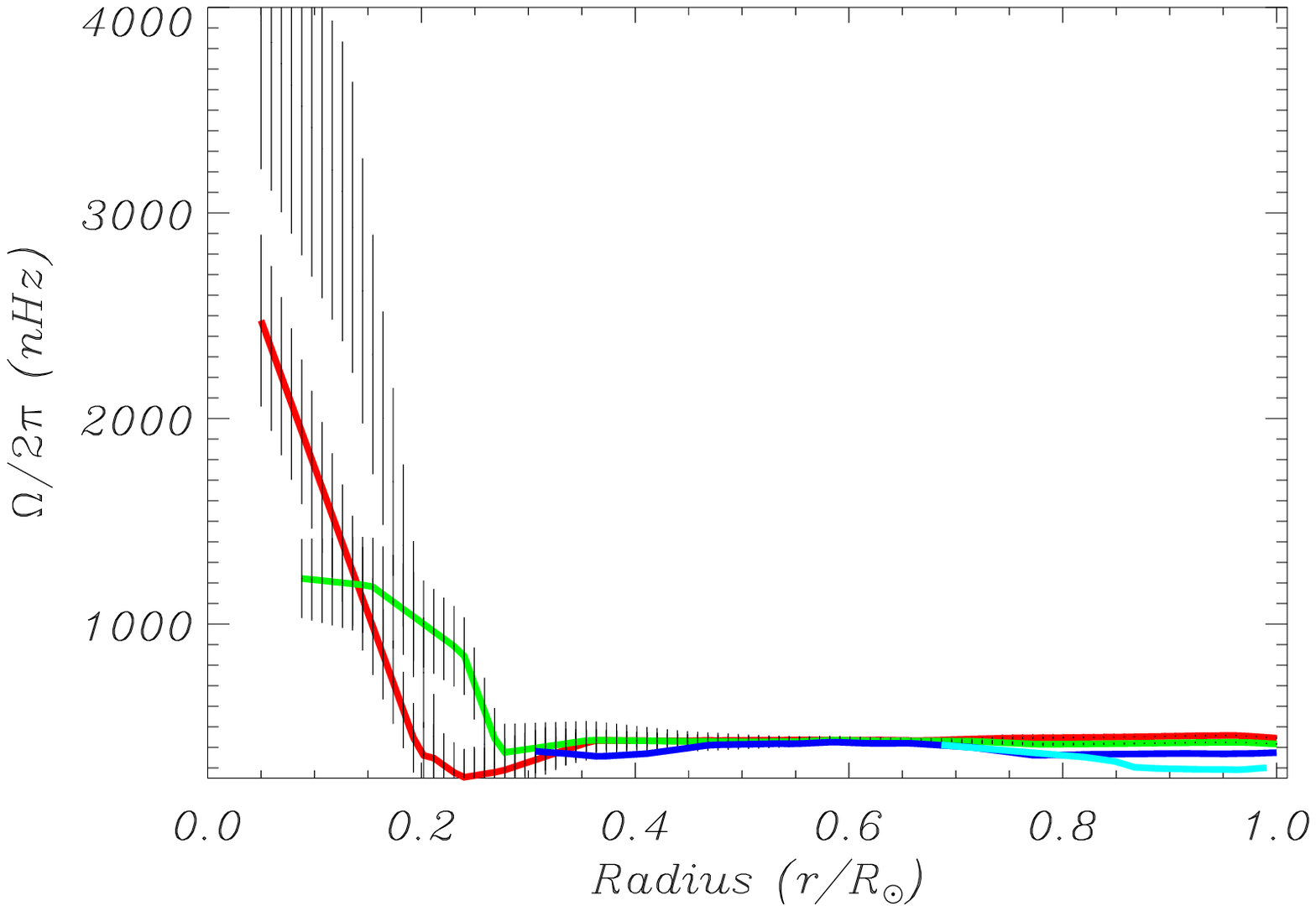}	

\caption{Inversions of the internal rotation rate of the Sun using 4608 days of MDI and GOLF data. On the left containing only p modes, on the right we have added the five candidate g modes of this work (with 40nHz error bars). The colors corresponds to the inversions at different latitudes: The equator (black), $20^{\circ}$ (red), $40^{\circ}$ (green), $60^{\circ}$ (blue) and $80^{\circ}$ (light blue).
\label{inv}}
\end{figure}

Once we have characterized the central frequency and the rotational splittings of these 5 dipole gravity modes, we can include them in an inversion of the internal rotation of the Sun. To do so, we use the p-mode frequencies and splittings extracted from 4608 days of MDI data [48] for the modes $\ell \ge 3$ and the ones from GOLF,  for the modes $\ell \le 2$. The method used for the inversion is a modified Regularized Least Squared methodology [49]. Figure~\ref{inv} (left) shows the inversion of the internal rotation without including the five g-mode candidates, while on the right we show the resultant inversion including those g-mode candidates. 

Firstly, we notice that the inversions worked fine even considering that we have put a small error bar of 40nHz in the rotational splittings of the g-mode candidate. Some differences are found at different latitudes which can be a consequence of the small sensitivity we have in the core due to the use of only a few dipole modes. Finally, the average rotation rate obtained is compatible with the average rotation rate previously inferred using the asymptotic periodicity. However, more work is needed to understand the sensitivity of the error bars in the inversion.

\section{Conclusions}
In this paper we have been able to identify the individual peaks generating the periodic signal found by [8] and interpreted as the asymptotic periodicity of the dipole gravity modes. The analysis of the collapsograms gave five possible detections in the frequency range between 60 and 140 $\mu$Hz with a rotational splittings in the range 850 to 950 nHz. When these candidate modes are used in the inversion of the rotational profile, we see that the rotation in the deep core reaches values close to 4000 nHz. However, more work should be done to better characterize these g-mode candidates as well as the determination of the error bars that are very important for the inversions. 

\ack
SoHO is a space mission of international cooperation between ESA and NASA. DS acknowledges the support of the grant PNAyA 2007-62650 from the Spanish National Research Plan, J.B. the support through the ANR SIROCO, and R.A.G and S.T. thanks the support form the CNES and the French Stellar Physics National Plan (PNPS). NCAR is supported by the National~Science~Foundation.

\end{document}